\documentstyle[12pt,epsf]{article}
\begin {document}
~\hspace*{8.0cm} ADP-AT-96-8, revised\\
~\hspace*{8.9cm} astro-ph/9608052\\
~\hspace*{8.9cm} Astroparticle Phys., in press\\[2cm]

\begin{center}
{\large \bf Photon-photon absorption above a molecular cloud torus in blazars
}\\[1cm]
R.J. Protheroe$^1$ and P.L. Biermann$^2$\\
$^1$Department of Physics and
Mathematical Physics\\ The University of Adelaide, Adelaide, Australia
5005\\$^2$Max-Planck Institut f\"{u}r Radioastronomie, Auf dem H\"{u}gel
69, D-53121 Bonn, Germany\\[2cm]
\end{center}

\begin{center}
\underline{Abstract}\\
\end{center}

Gamma rays have been observed from two blazars at TeV energies.  
One of these, Markarian 421, has been observed also at GeV energies and 
has roughly equal luminosity per decade at GeV and TeV energies.
Photon-photon pair production on the infrared
background radiation is expected to prevent observation above $\sim 1$ TeV.
However, the infrared background is not well known and it may be possible
to observe the nearest blazars up to energies somewhat below $\sim 100$ TeV
where absorption on the cosmic microwave background will give
a sharp cut-off.

Blazars are commonly believed to correspond to low power radio galaxies, seen
down along a relativistic jet; as such they are all expected to have the
nuclear activity encircled by a dusty molecular 
torus, which subtends an angle of 90 
degrees or more in width as seen from the central source.
Photon-photon pair production can also take place on the infrared radiation
produced at the AGN by this molecular torus and surrounding outer disk.
We calculate the optical depth for escaping $\gamma$-rays produced
near the central black hole and at various points along the jet axis 
for the case of blazars where the radiation is observed in a direction
closely aligned with the jet.

We find that the TeV emission site must be well above the top of the torus.
For example, if the torus has an inner radius of 0.1 pc and an
outer radius of 0.2 pc, then the emission site in Mrk~421 would have
be at least 0.25 pc above the upper surface of the torus, and if
Mrk~421 is observed above 50 TeV in the future, the emission site
would have to be at least 0.5 pc above the upper surface.
This has important implications for models of $\gamma$-ray emission in 
active galactic nuclei.

\section{Introduction}
The second  EGRET catalog of high-energy $\gamma$-ray sources
\cite{Thompson95} contains 40 high confidence identifications of
AGN, and a further 11 AGN identifications with lower confidence.
They show spectra where the luminosity is of similar order
of magnitude across the entire electromagnetic wavelength range,
from far infrared/mm wavelengths to GeV gamma photon energies, with
at times the gamma luminosity exceeding all other wavelengths
substantially.
The variability of blazars in all wavelengths, including the range
explored by EGRET on board the Compton Gamma Ray Observatory, 
is known to be extreme.
The variability both in the radio and in the gamma
range can reach time scales as short as hours.
Several of the AGN detected by EGRET have shown variability with
time scales of a few days for factors of $\sim 2$ changes in flux
\cite{Hartmann92,Hunter93,Kniffen93}.
One of the EGRET blazars, Markarian 421, has been observed at
TeV energies \cite{Punchetal,Mohanty,Kerrick95,Petry96}, 
as has Markarian 501  \cite{Quinn96}
which was not detected by EGRET.
At TeV energies Markarian 421 is extremely variable, and very recently
outbursts of TeV photons from Markarian 421 \cite{Mrk96flare}
have been observed with flux increases by a factor of 2 in one hour.
Such rapid variability indicates that the emission is produced
in the jet where relativistic effects can lead naturally to the
short timescales.
Furthermore, all of the active galactic nuclei (AGN) observed by EGRET are
blazars, objects in which the jet moving relativisticly is pointing
approximately towards the observer.
It is therefore natural to suppose that the $\gamma$-ray emission
is produced in the jet as a result of particle acceleration and
interaction with ambient medium, radiation or magnetic field.

There are two basic approaches to interpreting and modelling the
spectral energy distributions of these blazars.
The first approach involves the acceleration of electrons and subsequent
inverse Compton interactions of electrons with ambient radiation
to produce the $\gamma$-rays.
The ambient radiation may be radiation coming directly from
the accretion disk
\cite{MeliaKonigl89,DermerSchlickeiserMastichiadis92,DermerSchlickeiser93,
BednarekKirkMast96,ProtheroeStanev95},
or reprocessed by clouds or in an accretion disk corona
\cite{Blandford93,SikoraBegelmannRees93,Blandford95}.
These models have been quite successful in fitting the observational
data.

The second approach is to accelerate protons instead of, or as well as,
electrons.  Generally, the electron models have the emission region in the
jet rather closer to the central object than the models involving
proton acceleration.
In the proton  models, energetic protons interact with radiation by pion
photoproduction.
Again, this radiation may be reprocessed or direct accretion disk radiation
\cite{Protheroe96},
or may be produced locally, for example, by synchrotron radiation by
electrons accelerated along with the protons
\cite{MannheimBiermann92,Mannheim93a,Mannheim93b,Mannheim96}.
Pair synchrotron cascades initiated by photons and electrons resulting from
pion decay give rise to the emerging spectra, and this also leads to
quite acceptable fits to the observed spectra.
This second approach has the obvious advantage of leading to potentially
much higher photon energies, because protons have a much lower
synchrotron energy loss rate than electrons for a given magnetic environment.
In both classes of model, shock acceleration has been suggested
as the likely acceleration mechanism (see 
refs.~\cite{Drury,BlandfordEichler87,BerezhkoKrymsky88,JonesEllison91}
for reviews of shock acceleration), and
maximum energies were obtained and discussed in some detail in 
ref.~\cite{BS87}.

The observations clearly indicate that blazars are fairly low luminosity
radio galaxies seen along a line of sight rather close to the symmetry
axis of the bulk relativistic motion.
These radio galaxies are well known to enshroud the central source in a
dusty torus, which may cover a large fraction of $4 \pi$, as seen from the
central source.
The role of this dust torus has been the object of many investigations
\cite{jd1,jd2,jd3,jd4}, and has been discussed at some length in \cite{jd3},
where many earlier references are given 
(see also \cite{Barthel89a,Barthel89b}).  

Energetic photons may interact with low energy photons by photon photon 
pair production if the centre of momentum frame energy is above
threshold, i.e. $\sqrt{s} > 2 m_e c^2$.
Such interactions with the cosmic microwave background severely
limit the distance $\gamma$-rays above $\sim 100$ TeV can travel
from sources \cite{Gould,Jelley}.
At lower energies interactions on the infrared
and optical backgrounds limit the transparency at TeV energies 
(e.g. \cite{StdeJS,Pro93}).
Within the central regions of the AGN, interactions with photons 
directly from the 
accretion disk are important \cite{BeckerKafatos95,Bednarek96}
(See Bednarek \cite{Bednarek93} for an earlier discussion of $\gamma$-ray
escape from the radiation field of an accretion disk surrounding
a neutron star in an X--ray binary source).
In the present paper we wish to
present new calculations which give the photon photon opacity of gamma photons
produced in the jet, in the photon bath of the infrared torus which
surrounds the accretion disk.  
We will
demonstrate that the origin of the high energy gamma photons can only be above
the infrared torus by a factor of a few.  This excludes many of the published
models for $\gamma$-ray emission and variability.

\section{The basic geometry of the torus and calculation of the opacity}

From early on, there were attempts to unify and simplify the various 
classes of active galactic nuclei \cite{Scheuer79}.  
The most successful has been those models which unify the 
radio-loud AGN, where the  two main parameters are only the aspect 
angle and the total power of the
source \cite{Orr82}.   Attempts to include also the radio-weak AGN into such 
a scheme, have become more complicated \cite{Strittmatter80,jd1,Falcke96}. The 
aspect angle is the angle between  the line of sight to the observer, and 
the symmetry axis of the AGN itself, presumably the axis of the powerful  
jet, which in turn is usually taken  to be perpendicular to the compact 
accretion disk around a black hole.  The total power of the source is the
electromagnetic luminosity, integrated over all wavelengths (sometimes 
dominated by GeV $\gamma$-rays), plus the power flowing along the jet, 
in thermal
particles, in magnetic energy, in relativistic particles such as cosmic 
rays, and in kinetic luminosity, plus any other power channel, such as 
energetic particles coming out directly ({\it e.g.} neutrinos, neutrons, 
cosmic ray particles independent of the jet energetics).

It was recognized very early that the aspect angle plays a crucial role 
in our observations of AGN, on the one side with respect to the relativistic
boosting possible for small angles between jet and the line of
sight \cite{Scheuer79,Orr82,Antonucci85,Wills86}, and on the other side, 
with respect to absorption and scattering by surrounding material, nowadays
usually referred as the torus \cite{Mushotzky82,Lawrence82,Antonucci85}.

Since the far infrared emission of quasars was realized as arising from
molecular material by Chini {\it et al.} \cite{Chini89a,Chini89b} the 
far infrared emission of AGN has also become a favorite topic in these 
attempts to model the structure and emission region of AGN
\cite{Sanders89,PierKrolik,Niemeyer93}.  It now appears possible to begin
mapping at least some part of the molecular cloud distribution with
H$_2$O-maser lines such as in the galaxy NGC4258 and NGC1068
\cite{Barvainis95,Miyoshi95,Greenhill96}.

As a result we now have a fairly well established model that suggests
\cite{Antonucci93,Urry96} 
that the blazars are the low power end
of a distribution of intrinsic luminosities, seen under various aspect angles;
however, due to relativistic boosting, they belong to the most luminous
sources observed.  Key in such a picture is the inferred property of the
molecular cloud tori to cover a larger fraction of $4 \pi$ with decreasing
luminosity of the source \cite{jd3}.

The model by Pier \& Krolik \cite{PierKrolik} is a very good starting point 
to investigate the consequences of the radiation field given by the molecular
cloud torus, and we model the molecular torus following in the same way.
The torus is
of height $h$, symmetrical about the plane of reference, and centred on
the basic AGN source.
The cross section of the torus is approximated by a rectangle, and the
inner and outer radii are $a$ and $b$.
Typically, $a \sim 1$ pc, $b \sim 2 a$, and $h \sim a$ to $10a$.
For simplicity we model the torus with a single 
black body temperature of 1000 K \cite{PierKrolik}, and
surround the torus with a disk emitting black body 
radiation at a lower temperature
which decreases from 1000 K to 30 K between radius $r=b$
and $r= 100$ pc using a local power-law approximation, i.e.
$T(r) = 1000 (r/b)^{-\alpha}$~K, where $\alpha =1.52/\log(100 \; {\rm pc}/b)$.
Fig.~1 shows the location of the emission region in relation to the
torus and outer disk.
The figure also illustrates the interaction of a $\gamma$-ray emitted 
at point A
interacting at point B with a photon emitted from the upper surface of the
torus.

We then ask the following question:  What is the optical depth to
gamma-gamma opacity from pair creation of a gamma-photon in the bath
of the surrounding infrared light.
Using the top of the torus as our reference level, i.e.  $z=0$
corresponds to height $h/2$ above
the midplane in which the central source lies, we inject a $\gamma$-ray
on the axis at height $z=\ell$ (negative $\ell$
corresponds to injection inside the tunnel surrounded by the torus)
and calculate the optical depth out along the axis to infinity.

Consider a $\gamma$-ray photon of energy $E$ propagating out along the 
axis from $z =\ell$ to $z = \infty$ in the radiation field of the
upper surface of the torus.
The interaction probability depends on the angle, $\theta$,
between the directions of the $\gamma$-ray and infrared photon.
When the photon is at height $z$ above the surface, the maximum and
minimum values of $\cos \theta$ are

\begin{eqnarray}
(\cos \theta)_{\rm min} &=& z / (b^2 + z^2)^{1/2}, \\
(\cos \theta)_{\rm max} &=& z / (a^2 + z^2)^{1/2}.
\end{eqnarray}

The optical depth of the path from $z$ to $(z + dz)$ is

\begin{equation}
d \tau(E) = dz \int_{\varepsilon_{\rm th}}^\infty d \varepsilon n(\varepsilon)
\int_{(\cos \theta)_{\rm min}}^{(\cos \theta)_{\rm max}}
d \cos \theta {1 \over 2} (1 - \cos \theta) \sigma_{\gamma \gamma}(s)
\end{equation}

\noindent where

\begin{equation}
\varepsilon_{\rm th} = s_{\rm th} /2E(1 - \cos \theta_{\rm min}),
\end{equation}

\noindent $n(\varepsilon) d \varepsilon $ is the number density of photons of
energy $\varepsilon$ to $(\varepsilon+d \varepsilon)$ of black body
radiation at the temperature of the torus,

\begin{equation}
s = 2E \varepsilon (1 - \cos \theta)
\end{equation}

\noindent is the square of the centre of momentum frame energy,
$s_{\rm th} = (2 m_e c^2)^2$,
and
$\sigma_{\gamma \gamma}$ is the cross section for photon-photon
pair production \cite{Jauch} (see ref.~\cite{Prot92}, and references therein,
for a discussion of cascading in the radiation field along the
axis of a luminous disk).
After a change of variables, we can write

\begin{equation}
d \tau(E) = {dz \over 8 E^2}
\int_{\varepsilon_{\rm th}}^\infty d \varepsilon
{n(\varepsilon) \over \varepsilon^2}
\int_{s_{\rm 0}}^{s_{\rm max}} d s s \sigma_{\gamma \gamma}(s)
\end{equation}

\noindent where

\begin{eqnarray}
s_{\rm min} &=& 2E \varepsilon [ 1 - (\cos \theta)_{\rm max}],\\
s_{\rm max} &=& 2E \varepsilon [ 1 - (\cos \theta)_{\rm min}],
\end{eqnarray}

\noindent and $s_0$ is the larger of $s_{\rm min}$ and $s_{\rm th}$.
Integrating over $z$ from $\ell$ to $\infty$ we obtain the contribution to
the optical depth due to the top surface of the torus.
With slight modifications to the procedure described above, we can
easily obtain the contributions from the inner surface of the torus,
and the outer disk (for computational reasons this is split into
many annular rings each with a different temperature decreasing
appropriately with radius).

\section{The results}

In Fig.~2 we show the optical depth for $\gamma$-rays produced 
at the central point of symmetry, i.e. centre of the torus $\ell=-h/2$,
where by construction the AGN should be.
Results are given for $a = 1$ pc, $b \, = \, 2a$, for the cases
$h \, = \, a$ (bottom solid curve), $ 2a, \, ..., 10a$
(top solid curve), and because $\tau_{\gamma \gamma}$ scales with $a$
we show the optical depth in units of $\tau_{\gamma \gamma} /a$.
The dotted curve shows the contribution to $\tau_{\gamma \gamma}$
from $z > 0$.
These curves demonstrate already that for photon energies above
about 300 GeV the opacity is so large that no appreciable
radiation can emerge if the source of $\gamma$-rays is near the centre of the
torus.

Fig.~3 shows how far above the top of the torus we need to be
in order to be able to receive high energy photons.  These curves,
now from top to bottom, give the cases $\ell \, = \, -a, \, 0, \, a, ...
10a$ for $b \, = \, 2a$ and $a/h \, = \, 0.3$.  This means that this
sequence supplements the sequence in Fig. 2.  
Fig. 4 shows the height of the emission region such that the optical depth
is $\tau_{\gamma \gamma}=1$ 
for the case $b \, = \, 2a$ and $a/h \, = \, 0.3$ (as in Fig.~3),
and for a range of inner torus radius, $a$.
It is plotted in this way to show more clearly how high above the 
torus it is necessary for the emission region to be located 
for $\gamma$-rays of a particular energy to escape.
Here we see that
going far above the torus helps very clearly, with TeV photons
possible already at the top plane of the torus, and with a
level of $5a$ even the majority of 30 TeV photons would escape.

Fig. 5 shows how different surfaces of the torus contribute to 
the optical depth.
The case considered is the outermost case of Fig.~3,
namely $b \, = \, 2a$, $a/h \, = \, 0.3$, and $\ell \, = \, -a$;
this means that we are considering a source inside the torus, but
not all the way down to the central source.  From the top down
the curves first show the total, i.e. the top most curve from Fig.~3,
then the contribution from the inside surface of the torus for
the time while the presumed $\gamma$-ray is still inside the torus
(dot-dash line), next the contribution from the inside
surface of the torus for the time while the $\gamma$-ray is above the
top surface of the torus (long dashed line; note we always integrate the
optical depth along the path from the initial point to infinity),
then the contribution from the top surface the torus (dotted line),
and finally, the contribution from the outer disk outside the torus
(short dashed line).
It is clear that the contribution from inside the torus along the path
inside the torus itself is dominant.
It is also clear that the outer disk does not contribute significantly.

\section{Discussion}

The spectrum of $\gamma$-rays from Mrk~421 observed by the Whipple Observatory
\cite{Mohanty} extends well beyond 1 TeV and shows some 
evidence of a steepening
and possible cut-off at energies above 2 TeV.
We now discuss whether this cut-off is likely to be due to attenuation
on the intergalactic infrared background radiation or in the 
radiation field of the torus.
We show in Fig.~6 the observed spectrum together with an attempt by
Entel and Protheroe \cite{Entel95} to estimate the spectrum
emerging from the AGN using an inverse problem approach
(the allowed source spectrum would lie within the shaded region).
In doing this, they assumed $H_0=50$ km s$^{-1}$ Mpc$^{-1}$
and that the infrared background was as given in ref. \cite{Pro93}.
The infrared background is not well known and there have been several
attempts recently to estimate the infrared background by integrating
the emission expected from distant galaxies contributing to the
background \cite{DeZotti95,MadauPhinney96,Macminn96,Vaisanen96}.
However the spectrum
in ref. \cite{Pro93} is consistent with the most recent estimates
based on COBE data \cite{Puget96}.
Note that source spectrum obtained by Entel and Protheroe \cite{Entel95} 
is consistent with an $E^{-2}$ spectrum with no attenuation or cut-off.
Mrk~421 is a relatively low-luminosity blazar and has an infrared luminosity 
$\sim 4 \times 10^{44}$ erg s$^{-1}$ assuming a distance of $\sim 180$ pc
(see e.g. ref.~\cite{vonMontigny95a}).
Such an infrared luminosity and a temperature of the torus of 1000 K
would imply a torus radius of $\sim 0.1$ pc.
We show in Fig.~6 the effect on an $E^{-2}$ spectrum of $\gamma$-rays 
of attenuation on the radiation from the torus for the parameters used 
in Fig.~3 and $a=0.1$ pc.
To be consistent with the inferred source spectrum requires 
$\ell$ to be greater than 0.25~pc for the assumed parameters or, 
more generally, $\tau_{\gamma \gamma}<1$ up to $\sim 4$ TeV implying
an emission region well above the upper surface of the torus (see Fig.~4).
If we had assumed
that the observed spectrum was equal to the spectrum emerging from the AGN
(i.e. no intergalactic attenuation) we would require
$\tau_{\gamma \gamma}<1$ up to at least 2 TeV, and 
we would still conclude that the emission region must be 
well above the upper surface of the torus.

The results demonstrate that a photon of high energy cannot escape
the pair creation opacity in the bath of the far-infrared
radiation of the torus.  Quantitatively, the opacity is too high
for a source very close to the central source at 300 GeV, and requires
an initial point of departure far above the torus at energies such
as 30 TeV.

Therefore, any source which is observed to have TeV photons, requires
the origin of these photons to be above the torus. For standard
parameters, i.e. parsec scale tori, this means that any source
at such photon energies requires the TeV photons to originate
at several parsecs from the central source of activity.
We next discuss which blazar models may be consistent with this constraint.

There are various models in the literature which try to explain the ubiquitous
high energy $\gamma$-ray emission detected for many flat-radio-spectrum 
blazars. 
These models basically fall into two main classes:

\begin{itemize}

\item{}  In the first group of models, the $\gamma$-ray emission is due to an
inverse Compton process, using high energy electrons colliding with external
photons directly from an accretion disk or scattered by hot gas surrounding
the disk.  The electrons in turn, may be produced by various processes.  Such
models have been described by groups involving Begelman, Blandford, Dermer,
Schlickeiser, Sikora, and others {\it e.g.}
\cite{Sikora87,Sikora90,DS92,DS93,Blandford95,DS94,Boettcher96}.  

\item{}  In another group of models, the basic processes to produce the
$\gamma$-ray photons are either inverse-Compton scattering or inelastic
proton-photon scattering of locally produced photons; in their application 
to $\gamma$-ray blazars, synchrotron self-Compton models have been developed by
Maraschi, Ghisellini and Celotti \cite{Maraschi92},
Zdziarski and Krolik \cite{ZdzK93}, and 
Bloom \& Marscher \cite{Bloom96}, and a hadronic model by Mannheim and
collaborators {\it e.g.} \cite{MannheimBiermann92,MKB,MB92,Mannheim93a}.

\end{itemize}

Of the published models which attempt 
to explain the high energy $\gamma$-ray emission
from blazars, it appears that all models in the first class
above, with the likely exception of \cite{Blandford95}, fail the test of the
constraint described in the present paper, whereas the models in this second
class are consistent with the new constraint

However, it is premature to eliminate all the models in the first class above,
because the authors of the propositions can be trusted to reconstruct their
models, or at least a subset of them, to meet the constraints described here. 
The  basic argument used by many of these authors is that the time-variability
observed for the $\gamma$-ray emission strongly argues for a distance from the
emitting region to the central engine, which is small; it is this step in the
argument which requires the most severe change, with all its consequences.

Apparently, rapid $\gamma$-ray variability means that the jet has a very small
opening angle between the inner accretion disk and the top of the
circum-nuclear torus and that the $\gamma$-rays are produced where the jet 
leaves the torus. 

Finaly we note that in a very recent paper, Ommer, Westerhoff and Meyer 
\cite{OmmerWesterhoffMeyer96} have searched in the HEGRA data for 
$\gamma$-ray emission above 50 TeV from a superposition of 14 blazars 
with redshifts less than 0.062, and find weak evidence for enhanced 
$\gamma$-ray emission from the directions of the blazars.
If these observations are confirmed they will have important implications 
for the infrared background radiation field, and for models of 
$\gamma$-ray emission in blazars.
For example, from Fig.~4 we see that 50 TeV $\gamma$-rays would have to be 
emitted a distance of at least 5 inner torus radii above the upper surface 
of the torus.

\section{Acknowledgments}
This work was started at a pleasant workshop on high energy neutrino physics at
Aspen, organized by Tom Weiler; we wish to thank him and the Aspen Center for
Physics for the beautiful surroundings and the generous hospitality.  
We wish to thank Drs. W. Bednarek, H. Falcke, Q. Luo, and K. Mannheim for a
critical reading of some parts of this paper.
We thank the referee for helpful comments on the original manuscript.
This research is supported by a grant from the Australian Research Council.

\newpage

\begin{figure}[htb]
\centering{\ \epsfbox{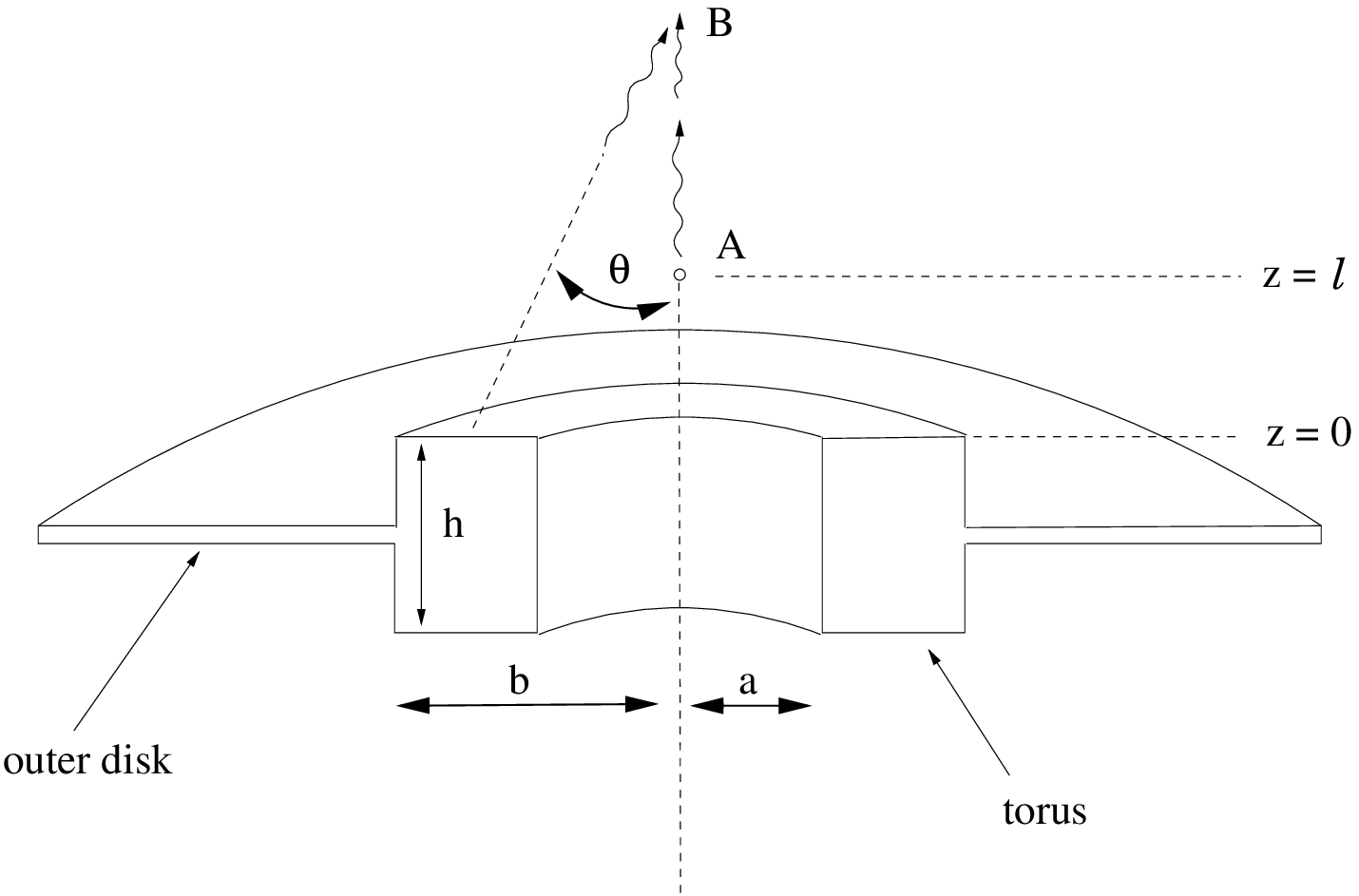}}
\caption{Geometry of the disk, torus and outer disk (not to scale)
used in the present work showing a $\gamma$-ray originating at point A
at height $z=\ell$ above the upper surface of the torus,
and travelling outwards along the jet axis to interact at point B
with a photon emitted from the upper surface of the torus.}
\end{figure}

\begin{figure}[htb]
\centering{\ \epsfbox{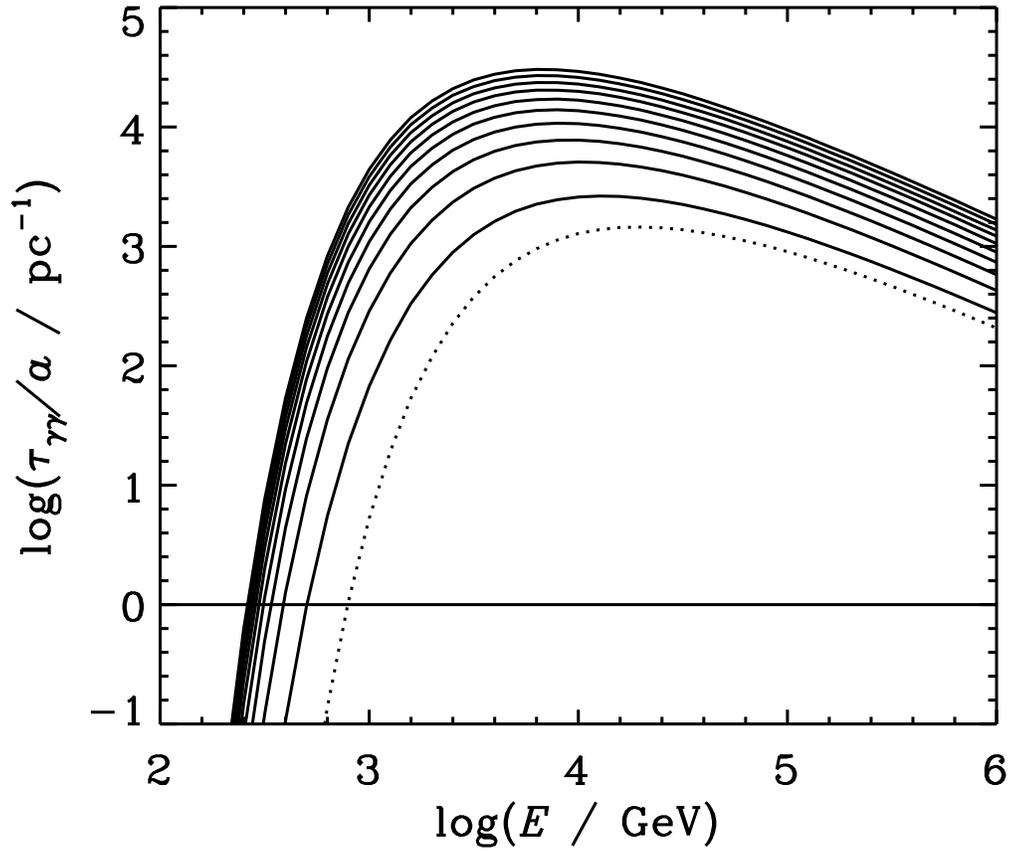}}
\caption{Optical depth for $\gamma$-rays produced at the
centre of the torus $\ell=-h/2$.
Results are given for $a = 1$ pc, $b \, = \, 2a$, for
$h \, = \, a$ (bottom solid curve), $ 2a, \, ..., 10a$
(top solid curve).
Dotted curve shows the contribution to $\tau_{\gamma \gamma}$
from $z > 0$.}
\end{figure}

\begin{figure}[htb]
\centering{\ \epsfbox{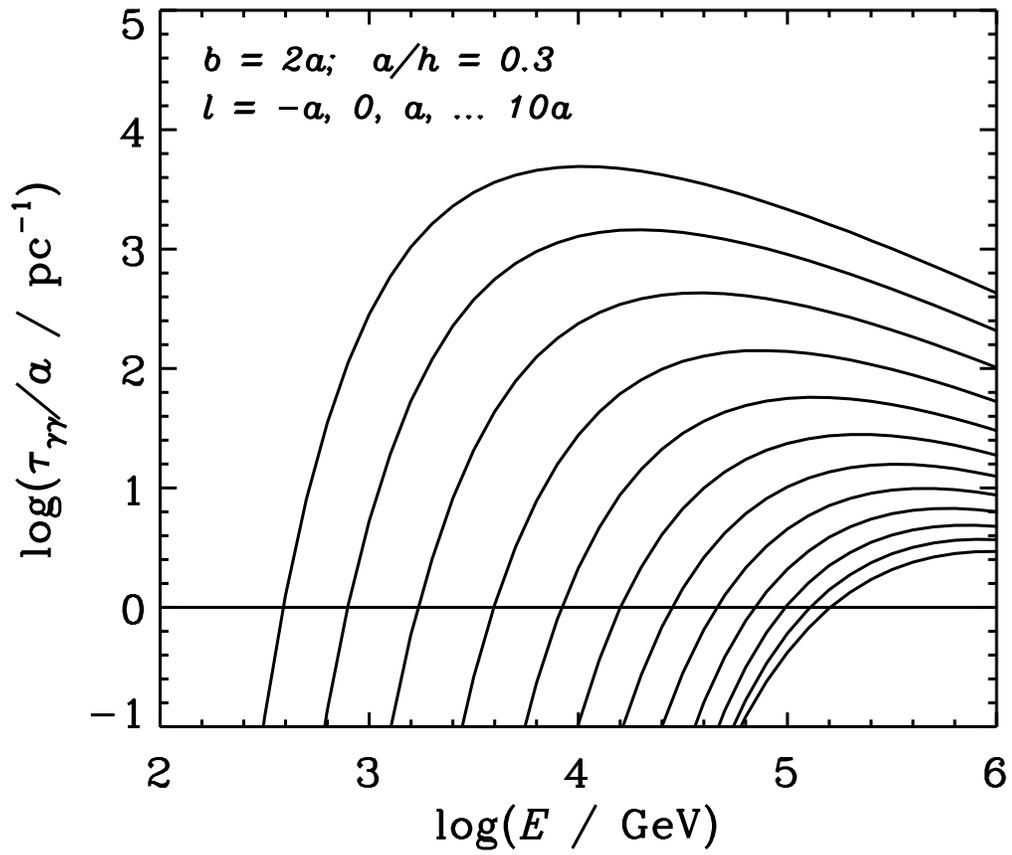}}
\caption{Optical depth for $\gamma$-rays produced at
$\ell \, = \, -a, \, 0, \, a, ...
10a$ (from left to right) for $b \, = \, 2a$ and $a/h \, = \, 0.3$.}
\end{figure}

\begin{figure}[htb]
\centering{\ \epsfbox{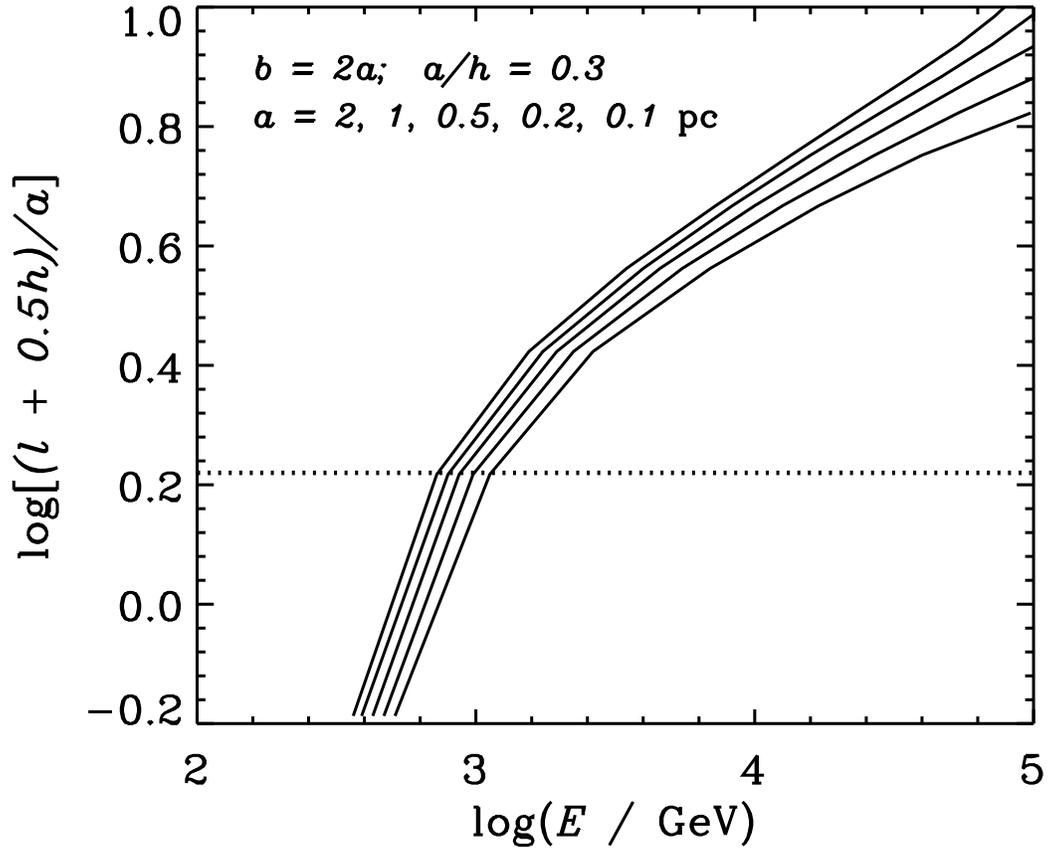}}
\caption{Height above the centre of the AGN $(\ell + 0.5 h)$
of the emission point (divided by $a$) where $\tau_{\gamma \gamma}=1$
for $a=2$, 1, 0.5, 0.2, and 0.1 pc (from top to bottom).
Dotted line corresponds to the level of the upper surface of the torus.}
\end{figure}

\begin{figure}[htb]
\centering{\ \epsfbox{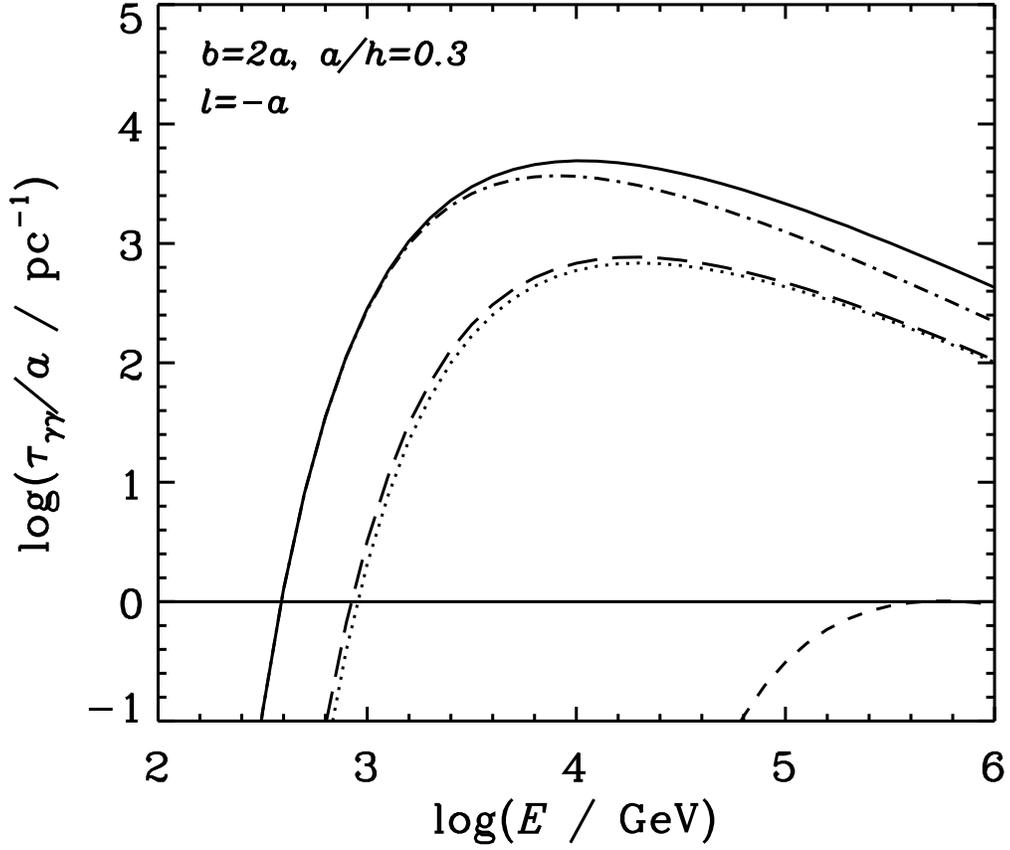}}
\caption{Optical depth for $\gamma$-rays produced at $\ell \, = \, -a$,
for $b \, = \, 2a$ and $a/h \, = \, 0.3$ showing how different
surfaces of the torus contribute: inside surface of the torus for
the time while the $\gamma$-ray is inside the torus
(dot-dash line), the contribution from the inside
surface of the torus for the time while the $\gamma$-ray is above the
top surface of the torus (long dashed line),
the contribution from the top surface the torus (dotted line),
and the contribution from the outer disk outside the torus
(short dashed line).}
\end{figure}

\begin{figure}[htb]
\centering{\ \epsfbox{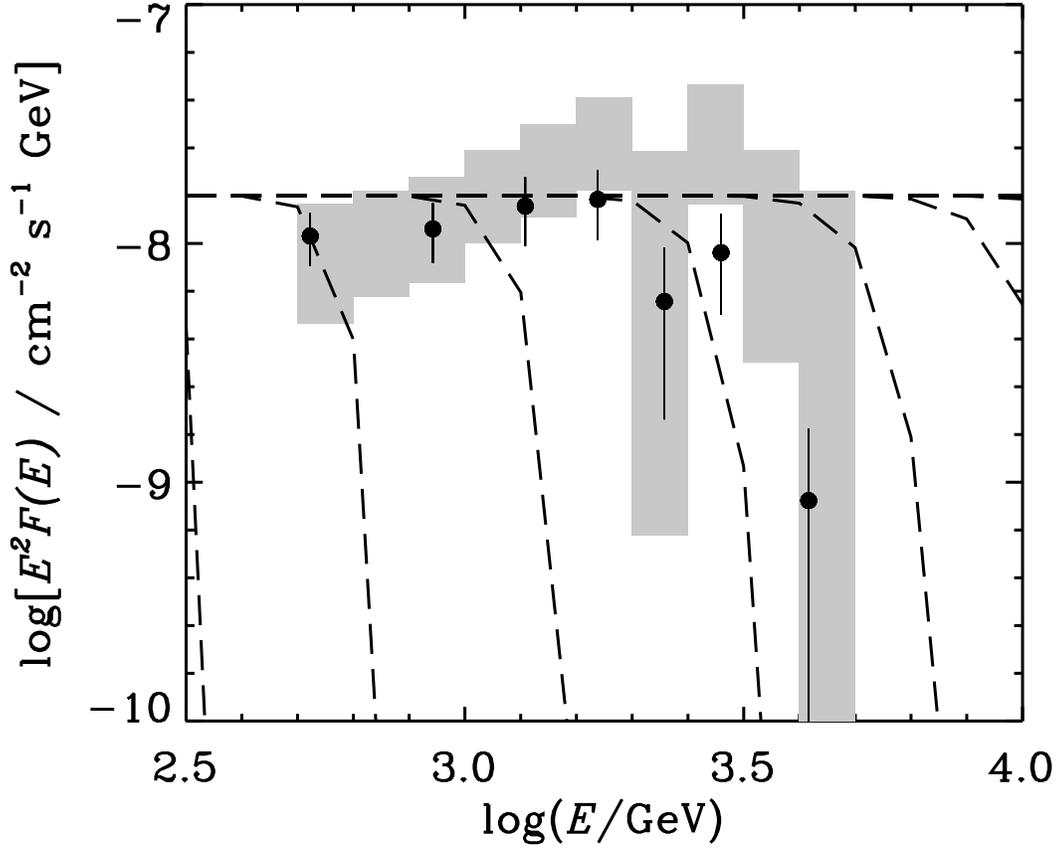}}
\caption{The observed VHE spectrum of Mrk 421 (ref. \protect 
\cite{Punchetal,Mohanty}) and the inferred range of source spectra (shaded) 
calculated in ref. \protect \cite{Entel95} 
assuming $H_0=50$ km s$^{-1}$ Mpc$^{-1}$
and the infrared background given in ref. \protect \cite{Pro93}.
Dashed curves show the effect on an $E^{-2}$ $\gamma$-ray spectrum 
of attenuation on the radiation from the torus for the parameters used 
in Fig.~3 and $a=0.1$ pc, i.e. $b \, = \, 0.2$ pc, $h \, = \, 0.33$ pc and 
$\ell \, = \, -0.1, \, 0, \, 0.1, ... 0.5$ pc (from left to right).}
\end{figure}

\end{document}